\input{psfig}
\documentclass[referee]{aa}
\usepackage{psfig}
\thesaurus{12.03.4; 12.12.1; 11.06.1}
\title{Density profiles of dark matter halos in an
improved Secondary Infall model}
\author{A. Del Popolo\inst{1,2} \and M. Gambera\inst{1,3} \and
E. Recami\inst{4,5,6} \and E. Spedicato\inst{2}}
\offprints{A. Del Popolo E-mail: adelpopolo@alpha4.ct.astro.it}
\institute{Istituto di Astronomia dell'Universit\`a di Catania, 
Viale A.Doria, 6 - I 95125 Catania, Italy  \and
Dipartimento di Matematica, Universit\`{a} Statale di Bergamo, Piazza Rosate 2 - I, 24129 Bergamo, Italy \and
Osservatorio Astrofisico di Catania and CNR-GNA, Viale A.Doria, 6 - I 95125 Catania, Italy  \and
Facolt\'a di Ingegneria, Universit\`{a} Statale di Bergamo, Viale G. Marconi 5, 24044 Dalmine, Bergamo, Italy \and
I.N.F.N., Sezione di Milano, Milano, Italy \and
C.C.S. and D.M.O./FEEC, State University at Campinas, S. P., Brazil
}
\titlerunning{Density profiles of dark matter halos}
\authorrunning{Del Popolo et al.}
\date{}
\begin{document}
\maketitle

\begin{abstract}
In this paper we calculate the density profiles of virialized halos both
in the
case of structure evolving hierarchically from a scale-free Gaussian $\delta
-$field having a power spectrum $P(k)\propto k^n$ in a $\Omega=1$
Universe and in the case of the CDM model, by using a modified version
of Hoffman \& Shaham's (1985) (hereafter HS) and Hoffman's (1988) model.
We suppose that the initial density contrast profile around
local maxima is given by the mean peak profile introduced by Bardeen et al.
(1986) (hereafter BBKS), and is not just proportional to the two-point
correlation function, as assumed by HS.
%and Hoffman
%(1988).
We show that the density profiles,
both for scale-free Universes and the CDM model,
are not power-laws but have
a logarithmic slope that increases from the inner halo to its outer parts.
%%and for $n \geq -1$ are
Both scale-free, for $n \geq -1$, and CDM density profiles are 
well approximated by
Navarro et al. (1995, 1996, 1997) profile.
%For all values of $n$ of interest, the slope of the
%virialized density profile is steeper than the value obtained by Hoffman \&
%Shaham (1985) in agreement with Sheth \& Jain's (1998) result.
The radius $a$, at which the slope $\alpha=-2$, is a function
of the mass of the halo and in the scale-free
models also of the spectral index $n$.
%Finally we show
%that also the density profile obtained by using a CDM spectrum and the same model
%is not a power low but is well fitted by the profile proposed by  Navarro et
%al. (1995, 1996, 1997).
\end{abstract}

\section{Introduction}

The collapse of perturbations onto local density maxima of the primordial
density field is likely to have played a key role in the formation of
galaxies and clusters of galaxies. The problem of the collapse
has been investigated from two points of view, namely that
of the statistical distribution of the formed objects (a question 
related to the biasing problem) (Kaiser 1984; Davis et al. 1985; BBKS) and
that of the structure of these objects and its dependence on the statistical properties of the
primordial density field (Gunn \& Gott 1972; Gunn 1977;
Filmore \& Goldreich 1984; Bertschinger 1985; West et al. 1987; HS;
Hoffman 1988; Efstathiou et al. 1988; 
Quinn et al. 1986; Warren et al. 1991; White \& Zaritsky 1992; Evrard
et al. 1993; Crone et al. 1994; Navarro et al. 1995, 1996, 1997;
Avila-Reese et al. 1998).\\
To overcome the problem of the excessively steep density profiles,
$\rho\propto r^{-4}$, obtained
in numerical experiments of simple gravitational collapse Gunn \& Gott (1972),
Gott (1975) and Gunn (1977) were able to produce shallower profiles, 
$\rho\propto r^{-2}$ through the ${\it secondary}$ ${\it infall}$ process.
Self-similar solutions were found by Fillmore \& Goldreich (1984) and
Bertschinger (1985), who found a profile of $\rho\propto r^{-2.25}$.
HS considered a scale-free initial
perturbation spectra, $P(k) \propto k^n$ and assumed that
local density extrema are the progenitors of cosmic structures and that the
density contrast profile around maxima is proportional to
the two-point correlation function. They thus showed that $\rho\propto {r^{-\alpha}}$ with
$\alpha=\frac{3(3+n)}{(4+n)}$, thus recovering Bertschinger's (1985) profile for $n=0$ and
$\Omega=1$. They also showed that, in an open Universe, the slopes of
the density profiles steepen with increasing values of $n$ and
with decreasing $\Omega$, reaching a profile
$\rho\propto r^{-4}$ for $\Omega \rightarrow 0$. Hoffman (1988) refined the
calculations of HS and made a detailed comparison
of the analytical predictions of the secondary infall model (hereafter SIM) with the
simulations by Quinn et al. (1986) and Quinn \& Zurek (1988). 
In spite of the high level of simplification of the SIM,  
and although
the formation of dark matter halos in the numerical simulations seems to
grow in mass by mergers with typically less massive halos,
%(chaotic) due to mergers
while the collapse in the SIM
is spherical symmetric and gentle, 
%The first studies
%these studies about the structure of objects identified as virialized
%in
those numerical simulations (Quinn et al. 
1986; Frenk et al. 1988)
%for the Gaussian CDM models
were in agreement
with the predictions of the SIM.
%An explanation to the good
The good results
given by the SIM in describing the formation of dark matter halos seem to be
due to the fact that in energy space the collapse is ordered and gentle,
differently from what seen in N-body simulations (Zaroubi et al. 1996). If things go
really in this way, it is possible that galaxies and clusters of galaxies
retain memory of their initial conditions.\\
A great effort has been dedicated to study
the role of initial conditions in shaping the final structure of the dark
matter halos; but, if on large scales
(evolution in the weakly non-linear regime)
the growing mode of the initial density fluctuations can be recovered if the
present velocity or density field is given
(Peebles 1989; Nusser \& Dekel 1992), on small scales shell crossing and
virialization contribute to make the situation less clear. 
To study the problem,
three-dimensional large-scale structure simulations were run with 
often conflicting results. While Quinn et al. (1986) and
Efstathiou et al. (1988)  found a connection
between the density profiles of collapsed objects and the initial fluctuation spectrum
for Einstein-de Sitter universes [in particular Efstathiou et al. (1988)
found density profiles steepening with increasing spectral index $n$],
West et al. (1987)
arrived at the opposite conclusion. In any case, the previous studies showed that the mass
density profiles steepened with decreasing $\Omega$. This result is in
agreement with that of HS.
More recent studies (Voglis et al. 1995; Zaroubi et al. 1996) showed a
correlation between the profiles and the final structures. Finally
Dubinski \& Carlberg (1991), Lemson (1995), Cole \& Lacey (1996), 
Navarro et al. (1996, 1997) and Moore et al. (1997) found that dark
matter halos do not follow a power law but
develop universal profile, a quite general profile
 for any scenario in which structures form due to hierarchical
clustering, characterized by a slope $\beta=\frac{d \ln \rho}{d ln r}=-1$ near
the halo center and $\beta=-3$ at large radii.
In that approach, density profiles can be fitted with a
one parameter functional form:
\begin{equation}
\frac{\rho(r)}{\rho_b}= \frac{\delta_n}{\frac{r}{a}\left(1+\frac{r}{a}\right)^2}
\label{eq:nfw}
\end{equation}
where $\rho_b$ is the background density and $ \delta_n$ is the 
overdensity [below we shall refer to Eq. (\ref{eq:nfw}) (Navarro et al. 1997) 
as the NFW profile].
The scale radius $a$, which defines the scale where the
profile shape changes from slope $\beta<-2$ to $\beta>-2$,
and the characteristic overdensity, $\delta_n$, are related because the
mean overdensity enclosed within the virial radius $r_{v}$ is $ \simeq 180$.
The scale radius and the central overdensity are directly related to the
formation time of a given halo (Navarro et al. 1997). The power spectrum
and the cosmological parameters only enter to determine the typical formation
epoch of a halo of a given mass, and thereby the dependence of the
characteristic radius on the total mass of the halo. Also these last results
are not universally accepted. Recently, Klypin et al. (1997) and Nusser \& Shet (1998)
challenged the claim that a one parameter functional form could fit the
density profiles using N-body simulations. Klypin et al. (1997), by using N-body
simulations of CDM-like models, showed that the scatter about a
one parameter fit is substantial, and that more than just one physical
parameter is needed to describe the structure of halo density profiles in agreement
with Nusser \& Sheth's (1998) conclusions.
In short, the question of whether galaxies and clusters
mass density profiles retain information on the initial conditions and
the evolutionary history that led to their formation
remains an open question. \\
In this paper, we introduce a modified version of
HS and Hoffman's (1988) models to study the shapes 
of the density profiles that result from the gravitational collapse. 
In particular, we relax the hypothesis that the initial density profile 
is proportional to the two-point correlation function, and use the density 
profiles given by BBKS.\\  
The plan of the paper is the following: in Sect. ~2 we show 
the reasons why the HS (1985)
model must be improved and how it can be improved. In Sect. ~3 we introduce
our model and in Sect. ~4 we show our results and finally in Sect. 5
we draw our conclusions.

\section{Limits of the SIM and Hoffman \& Shaham's approaches}

Most analytic work has focussed on studying the evolution of
isolated spherical systems (Gunn \& Gott 1972, Gott 1975; Gunn 1977;
Fillmore \& Goldreich 1984; Bertschinger 1985; HS;
Hoffman 1988; White \& Zaritsky 1992) because there is
no analytical technique to study the evolution of a system
starting from general initial conditions. It is known that, if the
initial density profile of a halo is a power law in radius,
$\delta(r) \propto r^{-m}$,
($\delta(r)$ is the mean density at a distance $r$ from a peak)
then the density
profile of the collapsed halo is also a power law:
\begin{equation}
\rho \propto r^{-\alpha}
\label{eq:alf}
\end{equation}
where
$\alpha=\frac{3 m}{1+m}$, if $m \geq 2$,
and $\alpha=-2$ if $m<2$ (Filmore \& Goldreich 1984; Bertschinger
1985). The restriction on $m$ can be relaxed if non-radial orbits
are permitted (White \& Zaritsky 1992). HS and 
Hoffman (1988) related this spherical solution to dark matter halos,
that form from initial gaussian density fields. In their paper
they noted that the density profiles of very high peaks for $r>r_{v}$
have the
same radial dependence as the correlation 
function of the initial field, $\delta \propto \xi \propto r^{-(3+n)}$.
Then HS set $m=3+n$, obtaining:
\begin{equation}
\alpha=\frac{3(3+n)}{(4+n)}
\label{eq:bet}
\end{equation}
for $n \geq -1$, and $\alpha=-2$ for $n <-1$. This last conclusion
($\alpha=-2$ for $n<-1$) is not a direct consequence of the HS
model, but it was an assumption made by the quoted authors
following the study of
self-similar gravitational collapse by Fillmore \& Goldreich (1984). In fact,
as reported by the same authors,
in deriving the relation between the density at maximum expansion and the final
one (see next section) HS assumed that each new shell that collapses
can be considered as a a small perturbation to the gravitational
field of the collapsed halo. This assumption breaks down for $n<-1$.
In reality, all the slopes of the power spectrum satisfying the condition
$n<-1$ constitute a big problem for SIM (Zaroubi et al. 1996). In fact,
as shown in Zaroubi's et al. (1996) simulations and Lokas' et al. (1996) paper, 
the slope $n=-1$ marks the transition between two different dynamical regimes.
In the cosmological context, such regimes correspond to a primordial
perturbation field whose power spectrum is dominated by the high wavenumber
modes, $n>-1$, or by the low wavenumber modes, $n<-1$, respectively. \\
In the regime $n<-1$, the
energy of the particles changes violently in time, and one expects at the end
a state of statistical equilibrium, similar to the violent relaxation
described by Lynden-Bell (1967). In this regime there is a strong dependence on the
boundary conditions and the dynamics strongly depends on the last collapsing
shell (Zaroubi et al. 1996). According to Lokas et al. (1996)  for $n<-1$
the fluctuations grow faster than in the linear theory. \\
In the $n>-1$ regime, the
dynamics of the collapsed shells is hardly affected by the ongoing collapse of
more distant shells. In this case, order is preserved in energy space and
according to Lokas et al. (1996) a slowdown in the growth rate of perturbations
is expected. In such a case, the SIM is expected to be a useful tool
for calculating the final virialized structure of collapsed halos. 
For these reasons in this paper we suppose that $-1 \leq n \leq 0$.  \\
%Hoffman \& Shaham's model is based on the assumption (as we see in next section)
%that the final relaxed density is proportional to the density at
%maximum expansion phase of each shell. Unfortunately, as admitted by the
%same authors, the assumption breaks down for $n<-1$
%(as previously stressed
%taking account of non-radial orbits Eq. (\ref{eq:bet}) holds also
%for $n \leq 1$). \\
Let us also add that Eq. (\ref{eq:bet}) is true only for regions outside the virial radius of
a dark matter halo (see Peebles 1974; Peebles \& Groth 1976; 
Davis \& Peebles 1977; Bonometto \& Lucchin 1978; Peebles 1980; Fry 1984).
In the inner regions of the halo, scaling arguments plus the stability
assumption tell us that $\xi(r) \propto r^{-\frac{3(3+n)}{(5+n)}}$ and
we expect a slope different from Eq. (\ref{eq:bet}). Syer \& White (1996)
found for the inner regions of the halo a profile
$\rho(r) \propto r^{-\frac{3(3+n)}{(5+n)}}$, coincident with the slope
of the correlation function. Nusser \& Sheth (1998) found
$\frac{3(3+n)}{(5+n)} \leq \alpha \leq \frac{3(3+n)}{(4+n)}$, while
Sheth \& Jain (1996) found $\alpha=\frac{3(4+n)}{(5+n)} $. 
In other words, HS's (1985) solution applies only to
the outer regions of collapsed halos and consequently the conclusion,
obtained from that model, 
that dark matter halos density profiles 
can be approximated by power-laws on their overall radius range
is not correct. It is then necessary to
introduce a model that can make predictions also on the inner parts of
halos. \\
Another problem of HS's work is the assumption that 
$\delta(r) \propto \xi(r) \propto r^{-(3+n)}$. This is true only for
very high peaks (see Ryden \& Gunn 1987). As the peak height decreases, the
peak profile becomes steeper than the correlation function (BBKS) and
consequently the final density profile becomes steeper. Moreover, according to
BBKS, the mean
peak profile depends on a sum involving the initial correlation function,
$\xi(r) \propto r^{-(5+n)}$,
and its Laplacian, ${\bf \bigtriangledown}^2 \xi(r) \propto r^{-(3+n)}$
(BBKS; Ryden \& Gunn 1987).
This means that there are at least two reasons why 
the density profile outside the virial radius must be steeper than
in HS's model:
\begin{itemize}
\item a) peaks that give origin to structures have
height $\nu=2,3$, and not $\nu \rightarrow \infty$; 
\item b) $\delta(r)$ depends on a sum of $\xi(r)$ and its Laplacian, and for
peaks $\nu=2,3$ is steeper than the correlation function.
\end{itemize}
Inside the virial radius HS's model cannot be used
to predict the virialized density profile because it was a priori
constructed to estimate density profiles at $r >  r_{v}$.\\
That model suffers also from another drawback: namely, the assumption that
the accreting matter is not clumpy, while in hierarchical scenarios
for structure formation halo grows in mass by merger with less massive
halos. As previously told according to Zaroubi et al. (1996) probably
this last problem is not so difficult it, compared with the other two. \\
In the following, we want to show that the predictive power of the
SIM is greatly improved when the problems reported at point a) and b) are
removed. These two problems are not problems of the SIM but only
problems introduced by the HS's implementation
of it.

\section{The model}

In the most promising cosmological scenarios, structure formation 
in the universe is generated through the growth and collapse 
of primeval density 
perturbations originated from quantum fluctuations (Guth \& Pi 1982; 
Hawking 1982; Starobinsky 1982; BBKS) in an inflationary 
phase of early Universe. 
The growth in time of small 
perturbations is due to gravitational instability. 
The statistics of 
density fluctuations originated in the inflationary era are Gaussian, and 
can be expressed entirely 
in terms of the power spectrum of the density fluctuations: 
\begin{equation}
P( k) = \langle |\delta_{{\bf k}}|^{2} \rangle 
\end{equation}
where 
\begin{equation}
\delta_{{\bf k}} =\int d^{3} k exp(-i {\bf k x}) \delta({\bf x})
\end{equation}
\begin{equation}
\delta({\bf x}) = \frac{ \rho ({\bf x}) - \rho_{b}}{ \rho_{b} }
\end{equation}
and $ \rho_{b} $ is the mean background density. 
In biased structure formation theory it is assumed that cosmic structures 
of linear scale $ R_f$ form around the peaks of the density field, 
$  \delta( {\bf x})$, smoothed on the same scale. \\
If we suppose we are sitting on a $\nu \sigma$ extremum in the the smoothed
density field, we have that:
\begin{equation}
\delta(0)= \nu \xi(0)^{1/2} = \nu \sigma
\end{equation}
together with:
\begin{equation}
{\bf \bigtriangledown} \delta(r) \mid_{r=0}= 0
\end{equation}
If the Laplacian of $\delta(r)$ is unspecified, that means that the extremum
may be a maximum or a minimum, 
the mean density at a distance $r$ from the peak is then:
\begin{equation}
\delta(r)= \nu \xi(r)/\xi(0)^{1/2}
\label{eq:delt}
\end{equation}
(Peebles 1984; HS). If we calculate the mean density
around maxima, as done by BBKS, by adding the constraint: 
%As shown by BBKS the mean density
%around maxima obtained by adding the constraint
\begin{equation}
{\bf \bigtriangledown}^2 \delta(r) \mid_{r=0}<0
\end{equation}
we find that the mean density around a peak is given by:
%is given by:
\begin{equation}
\langle \delta (r) \rangle =\frac{\nu \xi (r)}{\xi (0)^{1/2}}-\frac{\vartheta (\nu
\gamma ,\gamma )}{\gamma (1-\gamma ^2)}\left[ \gamma ^2\xi (r)+\frac{%
R_{\ast }^2}3\nabla ^2\xi(r) \right] \cdot \xi (0)^{-1/2} 
\label{eq:dens}
\end{equation}
(BBKS; Ryden \& Gunn 1987),
where $\nu $ is the height of a density peak, $\xi (r)$ is the two-point 
correlation function:
\begin{equation}
\xi(r)= \frac{1}{2 \pi^2 r} \int_0^{\infty} P(k) k \sin(k r) d k
\end{equation}
$\gamma $ and $R_{\ast}$ are two spectral parameters
given respectively by:
\begin{equation}
\gamma =\frac{\int k^4P(k)dk}{\left[ \int k^2P(k)dk\int k^6P(k)dk\right]
^{1/2}}
\end{equation}
\begin{equation}
R_{*}=\left[ \frac{3\int k^4P(k)dk}{\int k^6P(k)dk}\right] ^{1/2}
\end{equation}
while $ \vartheta (\gamma \nu ,\gamma )$ is: 
\begin{equation}
\theta (\nu \gamma ,\gamma )=\frac{3(1-\gamma ^2)+\left( 1.216-0.9\gamma
^4\right) \exp \left[ -\left( \frac \gamma 2\right) \left( \frac{\nu \gamma }%
2\right) ^2\right] }{\left[ 3\left( 1-\gamma ^2\right) +0.45+\left( \frac{%
\nu \gamma }2\right) ^2\right] ^{1/2}+\frac{\nu \gamma }2}
\label{eq:tet}
\end{equation}
In order to calculate $\delta(r)$ we need a power spectrum, $P(k)$.
In the following, we restrict our study to an Einstein-De Sitter ($\Omega=1$)
Universe with zero cosmological constant and scale-free density
perturbation spectrum $P(k)$
%both a scale-free power spectrum:
\begin{equation}
P(k)=A k^n
\end{equation}
with a spectral index in the range $-1 \leq n \leq 0$,
and also to a CDM Universe with 
spectrum given by BBKS:
\begin{eqnarray}
P(k) = Ak^{-1}\left[ \ln \left( 1+4.164k\right) \right] ^2 \nonumber\\
\left(192.9+1340k+1.599\times 10^5k^2+1.78\times 10^5k^3+3.995\times 
10^6k^4\right) ^{-1/2}
%e^{-k^2l^2/2}
\end{eqnarray}
%(Ryden \& Gunn 1987).
We normalized the spectrum by  
%The normalization constant $ A$ can be obtained,
imposing that the mass variance at $8h^{-1}Mpc$ is $\sigma _{8}=0.63$. 
In the case of a scale-free power spectrum, it
is easy to show that the two-point correlation
function can be expressed in terms of the confluent
hypergeometric function, $_1F_1$, and of the $\Gamma$ function as:
\begin{equation}
\xi(r)= \frac{1}{2 \pi^2 } \exp \left(-\frac{r^2}{4\beta }\right)\frac 1{2\beta ^{(n+3)/2}}\Gamma (\frac{n+3}%
2)_1F_1(\frac{-n}2;\frac 32;\frac{r^2}{4\beta })
\end{equation}
where $\beta=R_f^2/2$, being $R_f$
the filtering radius,
and the Laplacian of $\xi(r)$ as:
\begin{equation}
{\bf \bigtriangledown}^2 \xi(r)= -\frac{1}{2 \pi^2 } \exp \left(-\frac{r^2}{4\beta }\right)
\frac 1{2\beta ^{(n+5)/2}}\Gamma \left(\frac{n+5}%
2)_1F_1(\frac{-n-2}2;\frac 32;\frac{r^2}{4\beta }\right)
\end{equation}
and finally $\delta(r)$ is:
\begin{eqnarray}
\delta(r) & = & \left( \frac{\nu}{\xi(0)^{1/2}}-
\frac{\theta(\nu \gamma,\gamma)\gamma}{(1-\gamma^2)\xi(0)^{1/2}}\right)
\frac{1}{2 \pi^2 } \exp \left(-\frac{r^2}{4\beta }\right)\frac 1{2\beta ^{(n+3)/2}} \cdot \nonumber \\
 & & 
\cdot 
\Gamma \left(\frac{n+3}%
2\right) \cdot_1F_1 \left(\frac{-n}2;\frac 32;\frac{r^2}{4\beta }\right) 
+
\frac{\theta(\nu \gamma,\gamma)R_{\ast}^2}{3 \gamma(1-\gamma^2) \xi(0)^{1/2}}
\nonumber \\
 & &
\frac{1}{2 \pi^2 } \exp \left( -\frac{r^2}{4\beta }\right)\frac 1{2\beta ^{(n+5)/2}}
\Gamma \left( \frac{n+5}%
2\right)_1F_1\left( \frac{-n-2}2;\frac 32;\frac{r^2}{4\beta }\right)
\end{eqnarray}
In the case that $\nu$ is very large Eq. (\ref{eq:tet}) reduces to
\begin{equation}
\theta \rightarrow 3(1-\gamma^2)/(\nu \gamma)
\end{equation}
and the mean density is well approximated by Eq. (\ref{eq:delt}), which is
the approximation used by HS
to calculate $\delta(r)$. 
In reality, for peaks having $\nu=2,3,4$, the mean expected density profile
is different from
a profile proportional to the correlation function both for galaxies
and clusters of galaxies (see BBKS). For example for galaxies the CDM profile is
steeper than that proportional to $\xi(r)$ as shown by Ryden \& Gunn (1987)
with a discrepancy increasing with decreasing $\nu$. 
%The mean fractional density excess inside a given shell of radius $r$
%can be calculated as:
%\begin{equation}
%{\overline \delta}=\frac{3}{r^3} \int_0^r \delta(y)y^2 dy
%\end{equation}
As shown by Gunn \& Gott (1972), a bound mass shell having initial comoving
radius $x$
will expand to a maximum radius:
\begin{equation}
r_m=x/{\overline \delta(r)}
\label{eq:pee}
\end{equation}
where
the mean fractional density excess inside the shell, as measured at
current epoch $t_0$, assuming linear growth, 
can be calculated as:
\begin{equation}
{\overline \delta}=\frac{3}{r^3} \int_0^r \delta(y)y^2 dy
\end{equation}
At initial time $t_i$ and for a Universe with density parameter $\Omega_i$,
a more general form of
Eq. (\ref{eq:pee}) (Peebles 1980) is :
\begin{equation}
r_m=r_i\frac{1+{\overline \delta_i}}{{\overline \delta_i}-(\Omega_i^{-1}-1)}
\end{equation}
The last equation must be regarded as the main essence of the SIM. It tells
us that the final time averaged radius of a given Lagrangian shell
does scale with its initial radius.
Expressing the scaling of the final radius, $r$, with the initial one
by relating $r$ to the turn around radius, $r_m$, it is possible to write:
\begin{equation}
r=Fr_m
\end{equation}
where $F$ is a costant that depends on $\alpha$:
\begin{equation}
F = F(\alpha) = 0.186+0.156 \alpha+0.013 \alpha^2+0.017 \alpha^3-0.0045 \alpha^4+0.0032 \alpha^5
\end{equation}
(Zaroubi et al. 1996).
%where
%the mean fractional density excess inside a given shell of radius $r$
%can be calculated as:
%\begin{equation}
%{\overline \delta}=\frac{3}{r^3} \int_0^r \delta(y)y^2 dy
%\end{equation}
If energy is conserved, 
%$r_{vir}$ is $r_m/2$ (Hoffman 1988)
then the shape of the density profile at maximum of expansion is
conserved after the virialization,
and
%the relaxed density profile
is given by (Peebles 1980; HS; White \& Zaritsky 1992):
\begin{equation}
\rho(r)=\rho_i \left( \frac{r_i}{r} \right)^2 \frac{d r_i}{dr}
\end{equation}
The density profile is a function of three parameters: the spectral index
$n$, the density parameter $\Omega$, and the height of the density peak,
$\nu$. In the limit $\nu>>1$, the overdensity $\delta(r)$ is proportional
to the two-point correlation function and the density profile is
a function of $n$ and $\Omega$ only, and then the expected profile is
that by HS. 

\section{Results and discussion}

By using the model introduced in the previous section we have studied the
density profiles of halos in scale-free universes with
$-1 \leq n \leq 0$, and for a CDM model characterized by a BBKS spectrum. As
previously quoted, the chosen range of $n$ is dictated by the limits of
the SIM and by the values of $n$ interesting in the cosmological
context.
The results of our calculations are shown in Fig. $1-6$. \\
In Fig. 1, 2, 3
we have calculated the density profiles of halos in a scale-free
universe with $n=-1$, $n=0$, and in a CDM model. 
\begin{figure}[ht]
%\picplace{2.0cm}
\psfig{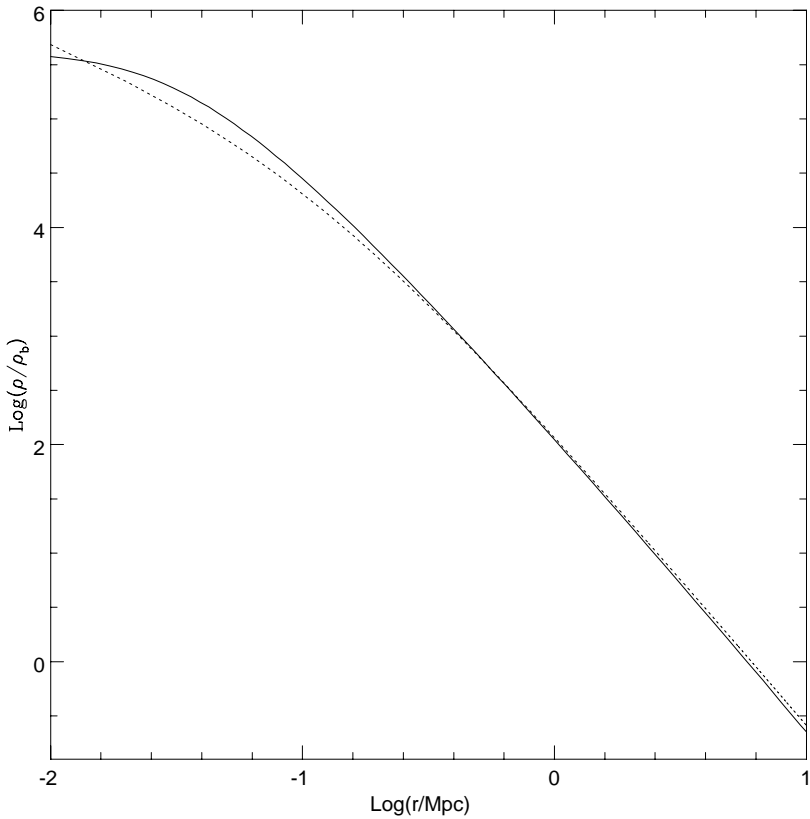}
\caption[]{Density profile for a scale-free spectrum with $n=-1$
(solid line) and the NFW fit (dashed line).
The mass of the halo is $\sim 2 \times 10^{15} M_{\odot}$.}
\label{Fig. 1}
\end{figure}
In Fig. 1 we show the density profile for $n=-1$. The dashed line is the
NFW fit, while the solid line represents the
profile obtained from our model. The NFW profile fits well the density
profile, except in the inner part.
%where the slope of our model is
%$\simeq -0.95$.
In the case $n=0$, see Fig. 2, the NFW (dashed line)
gives a good fit to the the density profile (solid line) also in the inner part of
the density profile.
%%As expected,
%%the density of the halo near the center is larger
%%with respect to the case $n=-1$ because halos form earlier in models
%%with larger $n$ and then have denser cores.
The situation for the halo obtained from the CDM spectrum, smoothed on
clusters scales $R_f=5h^{-1} Mpc$, 
(see Fig. 3) is similar to that of the case $n=-1$. This slope is in fact
similar to that of the standard CDM power spectrum on cluster scales. \\
In Fig. 4 we plot the slope of the density profile for several
values of $n$ and $\nu$. The fundamental aim of the picture is
to show how for small values of $\nu$ ($\nu=2,3$), out from
the inner region, the slope is larger than HF result.
We began by finding the HS solution, that was recovered as expected
in the limit
$\nu>>1$ or equivalently 
$\delta(r) \simeq \nu \xi(r)/\xi(0)^{1/2}$. This solution 
is represented by the short-dashed line which 
coincides with the result by HS, namely 
$\alpha=\frac{3(3+n)}{(4+n)}$, indicating
an increase in the slope $\alpha$ with increasing $n$.
Moreover the value of $\alpha$ is independent on the radius chosen to
compute the slope which means that halos are described by pure power-laws,
as described in the HF.
\begin{figure}[ht]
%\picplace{2.0cm}
\psfig{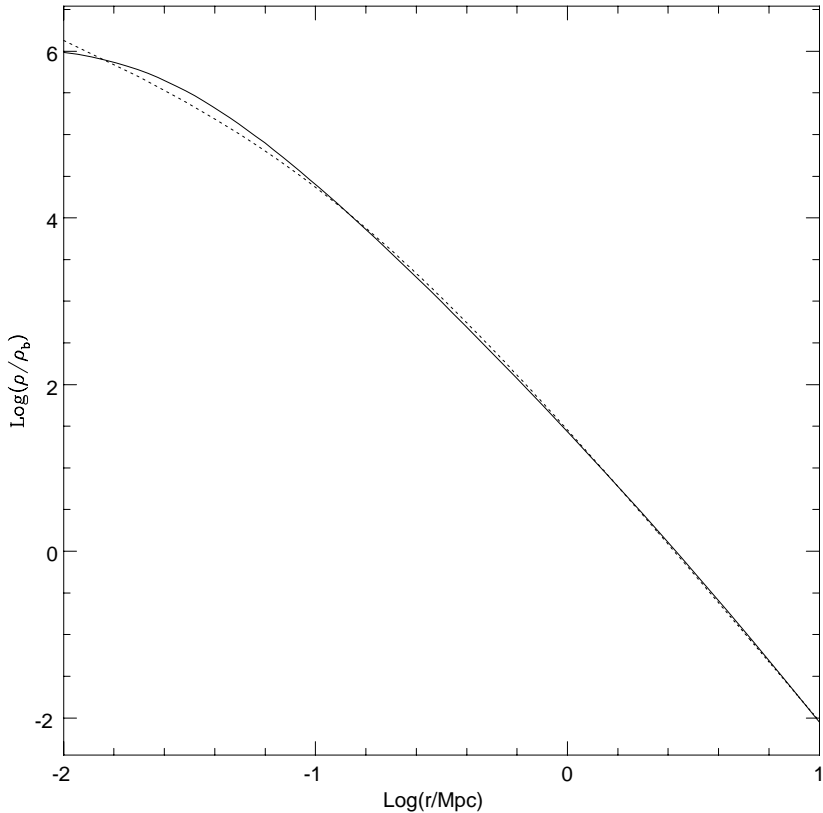}
\caption[]{As in Fig. 1, but now $n=0$. The mass of the halo
is $\sim 0.5 \times 10^{15} M_{\odot}$}
\label{Fig. 2}
\end{figure}
Because of the rarity of extremely high peaks, most galaxies and clusters
will form from peaks of height 2 or 3 $\sigma$ (BBKS; Ryden \& Gunn 1987): so
we repeated the calculation of $\alpha$ for these values. 
If we choose a value 
of $\nu=3$, the logarithmic slope of the density profile, calculated
at $1 h^{-1} Mpc$ (we calculated
the slope at a fixed distance because the density profiles are not
power-laws)
is steeper for all values of
$n$ (solid line) than that obtained by HS, and it is well approximated
by Shet \& Jain (1996) (dotted line), $\alpha=\frac{3(4+n)}{(5+n)}$,
obtained using
stable clustering and neglecting halo-halo correlations. At the same time
the dependence of $\alpha$ on $n$ is weaker than that shown by HS.
We compared our result to that of Shet \& Jain (1996)
%is: \\
%1) this model is one of the first analytical model showing the limits
%of HS model;\\
because their simple analytical
formalism describes reasonably well NFW profiles
on scales $r \ge 0.1 r_{\rm v}$
(at least till to $1 h^{-1}$ Mpc).
The choice of calculating the slope, $\alpha$, at $1 h^{-1}$ Mpc
is suggested from
the consideration that 
%1)
at that radius NFW profiles are well fit by Shet \& Jain (1996) model.\\
%2) density profiles described by our model (excluding the case $\nu>>1$)
%are not power laws
Obviously changing the radius at which the slope is calculted
this reflects on the relation $\alpha$-n, because the
density profiles described by our model (excluding the case $\nu>>1$)
are not power laws. For values larger than $1 h^{-1}$ Mpc, the value of the slope
is obviously larger. In any case, we have to remember that the comparison
with Shet \& Jain (1996) model is displayed only to show that density
profiles have larger slopes than those found by HF.\\
\begin{figure}[ht]
%\picplace{2.0cm}
\psfig{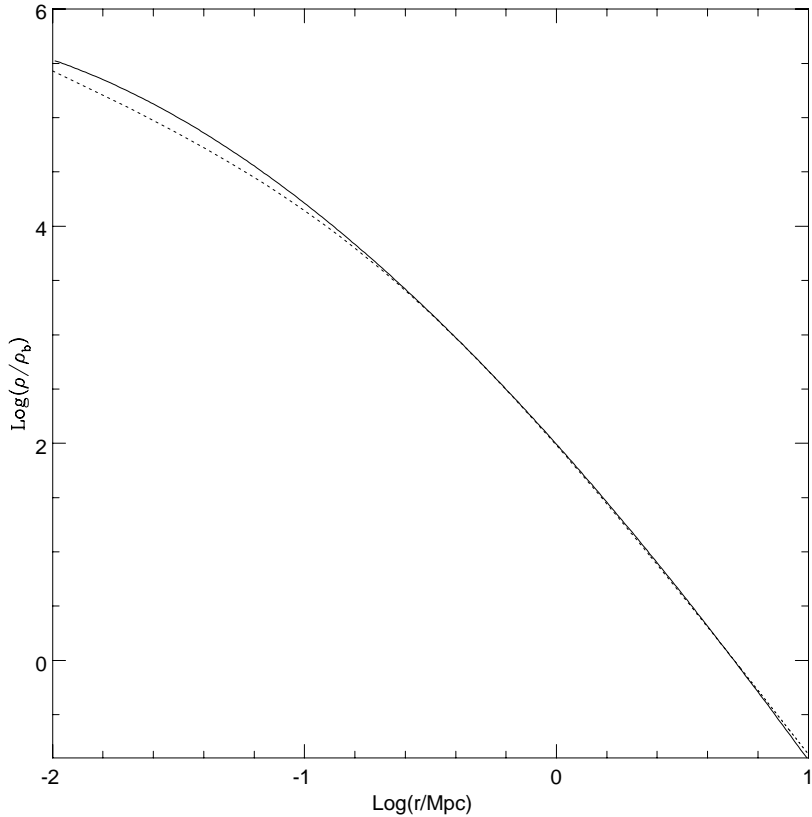}
\caption[]{Density profiles for a CDM spectrum smoothed on a scale
$R_f=5 h^{-1} Mpc$ 
(solid line) and the NFW fit (dashed line).}
\label{Fig. 3}
\end{figure}
For $\nu=2$ the slope is even steeper than the previous case (long-dashed line)
and it is well approximated by Crone's et al. (1994) result, which
is also consistent with the results by Navarro et al. (1997) (see their Fig. 13). As
in the previous case the dependence of $\alpha$ on $n$ is weaker with
respect to that shown in HS.
%In other words
More massive halos $\nu=3$ have flatter density profiles than less
massive ones in agreement with Tormen et al. (1997). \\
In Fig. 5 and Fig. 6 we plot $a/r_{\rm v}$, the variation of the ratio of
the scale parameter, $a$ and the virial radius, $r_{\rm v}$, 
versus
$M/M_{\ast}$ for a scale-free spectrum and a CDM spectrum, respectively.
We remember that according to Navarro et. al (1996, 1997), $a$ is linked to a dimensionless
"concentration" parameter, $c$, by the relation $a=\frac{r_{\rm v}}{c}$ and
the
parameter $c$ is linked to the characteristic density, $\delta_{\rm n}$,
by the relation:
\begin{equation}
\delta_{\rm n}=\frac{200}{3} \frac{c^3}{\ln(1+c)-c/(1+c)}
\end{equation}
The behaviour of the characteristic density of a halo,
increasing towards lower masses in all
the cosmological models, supports the idea that the
$M_{\rm v}$-$\delta_{\rm n}$ relation is a direct result of the higher
redshift of collapse of less massive systems. For scale-free models this
implies $\delta_{\rm n} \propto M^{-(n+3)/2}$ (same scaling relating
$M_{\ast}$ and the mean cosmic density at a fixed redshift z). \\
In the scale-free case, masses are normalized by the characteristic mass $M_{\ast}$,
which is defined at a time $t$ as the linear mass on the scale
currently reaching the non-linear regime:
\begin{equation}
M_{\ast}(t)=\frac{4 \pi}{3}R_{\ast}^3 \rho_b(t)
\end{equation}
where the scale $R_{\ast}$ is such that the linear density contrast on this
scale is $\delta(R_{\ast})=1.69$. Once known that the mass variance,
$\sigma_M$,
for a power
spectrum $P(k) \propto k^n$ is given by $\sigma_M \propto R^{-(3+n)}$
and remembering our normalization $\sigma_M(8 h^{-1} Mpc)=0.63$, the value of
$M_{\ast}$ for $n=-1$ results to be $M_{\ast}=6 \times 10^{13} M_{\odot}$.
\begin{figure}[ht]
%\picplace{2.0cm}
\psfig{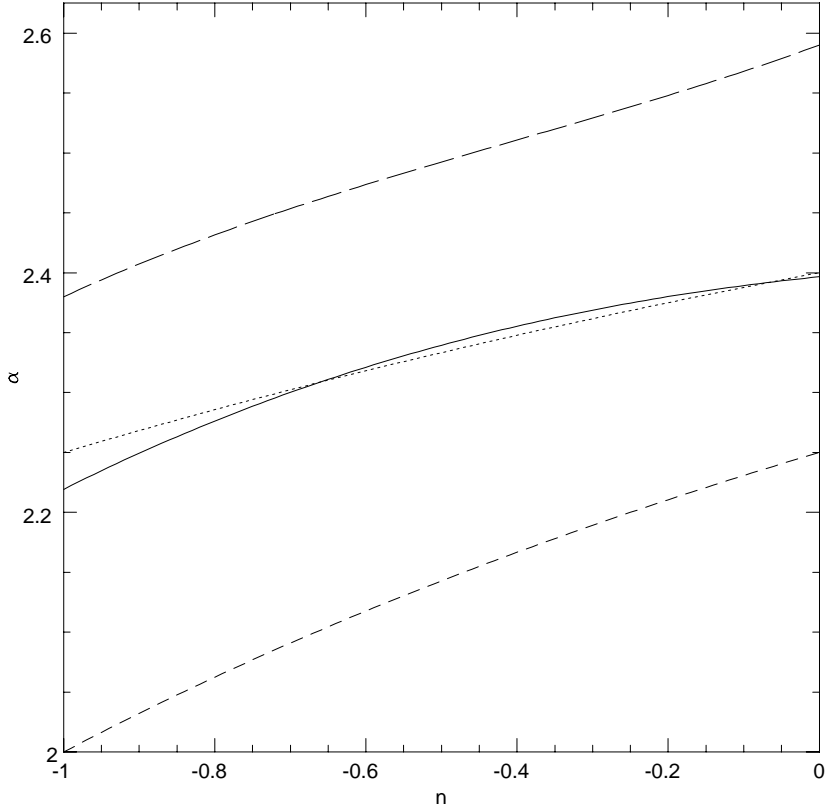}
\caption[]{The slope $\alpha$ of density profiles as a function of the spectral
index $n$ and $\nu$. The short-dashed line represents $\alpha$ in the
limit $\nu>>1$.
It 
coincides with the HS result. The solid line represents the logarithmic slope
for $\nu=3$, while the dotted line is Shet \& Jain's (1996) result. The
long-dashed line represents $\alpha$ for $\nu=2$.}
\label{Fig. 4}
\end{figure}
In the CDM case, the normalizazion mass, $M_{\ast}$, is obtained imposing
the condition $\sigma_{0}[M_{\ast}(z)]=1.69 (1+z)$. For our adopted
normalization of the CDM power spectrum,
$M_{\ast}=3 \times 10^{13} M_{\odot}$, at $z=0$.\\
In the scale-free case, Fig. 5, the value of the
%%dimensionless
scale radius $a$ correlates strongly
with halo mass and with spectral index $n$.
The
solid line represents $a$ for $n=-1$. As shown in the figure,
more massive halos have
a larger scale radius $a$, or equivalently less massive
halos are more concentrated. The dotted line shows $a$ for $n=0$.
Also for this value of $n$ more massive halos are less centrally
concentrated. Finally from Fig. 5 we also see that
in models with more small-scale power (or equivalently
larger values of $n$) the haloes tend to have denser cores.
These results were expected because halos with mass $M<<M_{\ast}$
form much earlier than haloes with $M>>M_{\ast}$ and then are more
centrally concentrated. Moreover, for a fixed value of $M/M_{\ast}$,
haloes form earlier in models with larger values of $n$ and then have
denser cores. This result is in qualitative agreement with those by 
Navarro et al. (1997),
Cole \& Lacey (1996),  Tormen et al. (1997).
The filled squares and the filled exagons, in Fig. 5,
represents $a/r_v$ for $n=-1$ and $n=0$ respectively obtained
by Navarro et al. (1997) in the case $f=0.01$ (see their paper for
a definition of this parameter) which give the best
fit to the results of their simulations.\\
The virial radius $r_v$ is
obtained by using Navarro et al. (1997) equation:
\begin{equation}
r_v=1.63 \times 10^{-2} \left(\frac{M}{h^{-1} M_{\odot}}\right)^{1/3}
\left(\frac{\Omega_0}{\Omega(z)}\right)^{-1/3} (1+z)^{-1} h^{-1} kpc
\end{equation}
where $\Omega_0$ is the actual value of the density parameter and $h=0.5$.
The virial radius determines the mass of the halo through:
\begin{equation}
M_v=200 \rho_b \frac{4 \pi}{3} r_v^{3}
\end{equation}
In the case $n=-1$, our model gives less concentrated halos till
$M \simeq 10 M_{\ast}$ and after this value the tendence is reversed.
Maximum deviations of $\sim 2.5$,
between Navarro et al (1997) data and our model, are
found in the low mass domain, $<0.25 M_{\ast}$. The situation is
similar to that described by Tormen et al. (1997) in his
{\it most relaxed} dynamical configuration (Fig. 16 second row in the
left panel).
In the case $n=0$, our model gives halos sligthly more concentrated
in the overall studied mass range. In this case the discrepancies reach
values of $\sim 3$.\\
\begin{figure}[ht]
%\picplace{2.0cm}
%%%%%%\psfig{file=fig2.ps,width=12.0cm}
\psfig{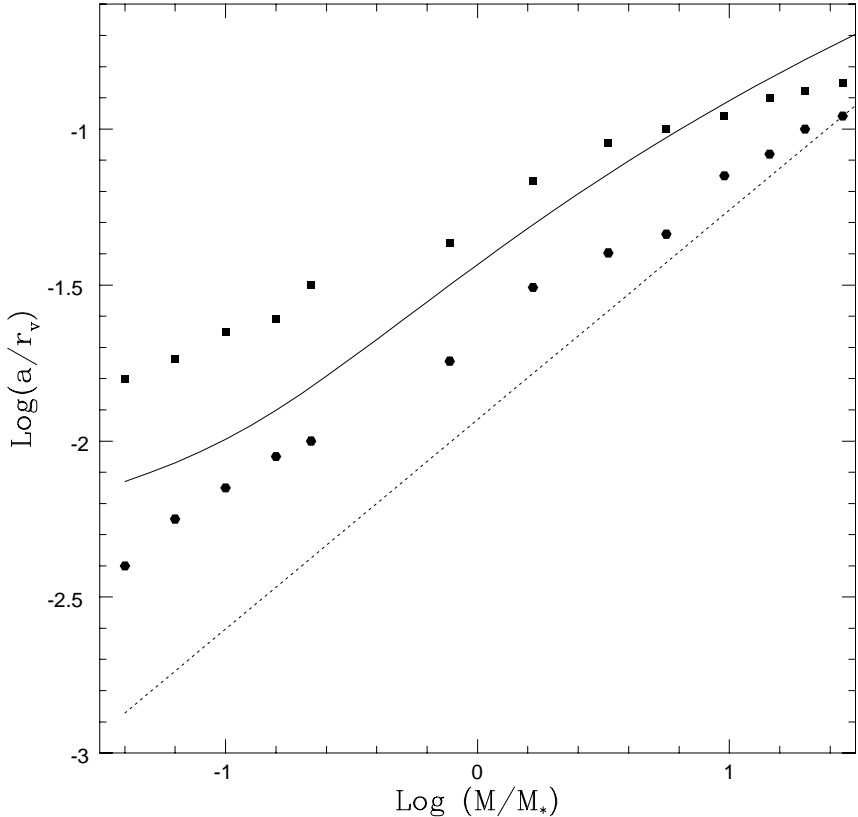}
\caption[]{Trend of the scale radius $a$ versus the mass of the halos in
the case $n=-1$ (solid line) and $n=0$ (dotted line).
The filled squares and the filled exagons
represents $a/r_v$, for $n=-1$ and $n=0$ respectively, obtained
by Navarro et al. (1997) in the case $f=0.01$.
}
\label{Fig. 5}
\end{figure}
Also in the CDM case, Fig. 6, the value of the
scale radius $a$ correlates strongly
with halo mass.
The
solid line represents $a$ for our model and the filled exagons the N-body
results of Navarro et al. (1996,1997). As for scale-free spectrum,
large halos are significantly less concentrated than small ones.
This trend, $a$ decreasing with decreasing $M$, can be explained
in terms of the formation times of halos (Navarro et al. (1996,1997)).
The dependence of concentration on mass is weak: there is only a change
of a factor of $\sim 4$ while $M$ varies by 4 orders of magnitude.
In this case the discrepancies between data and model is $\sim 2$. \\
The result obtained is a remarkable improvement of the SIM being it able
to reproduce almost all the prediction of N-body simulations with 
discrepancies of the same magnitude of those shown in Tormen et al. (1997)
(this for the case $n=-1$) and surely much smaller than those
found in Cole \& Lacey (1996).
\begin{figure}[ht]
%\picplace{2.0cm}
%%%%%%\psfig{file=fig2.ps,width=12.0cm}
\psfig{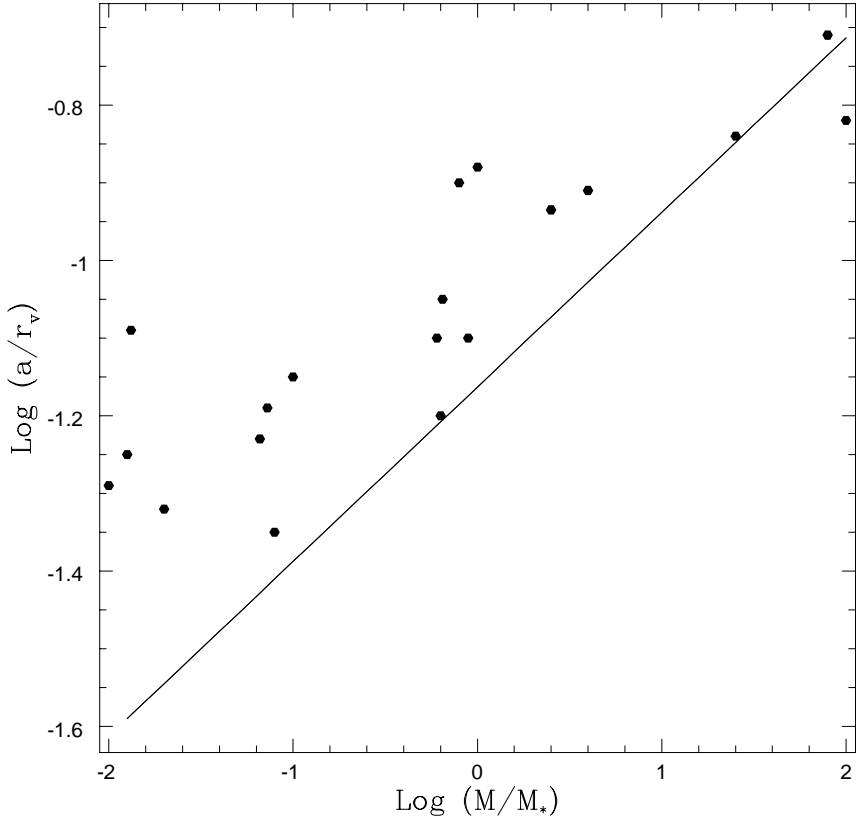}
\caption[]{Trend of the scale radius $a$ (solid line)
versus the mass of the halos in
the case of a CDM smoothed on a scale
$R_f=5 h^{-1} Mpc$ and with normalization $\sigma_8=0.63$. 
The filled exagons
represents $a/r_v$, obtained
by Navarro et al. (1996,1997).
}
\label{Fig. 6}
\end{figure}
Our model is based on spherical simmetry, and as we
previously stressed, halos accretion does not happen in
spherical shells but by aggregation of subclumps of matter which
have already collapsed. In other words it seems that the halos structures
does not depend crucially on hierarchical merging, in agreement with
Huss et al. (1998). The SIM seems to have more predictive
power than that till now conferred to it. 
%, remembering that the model is just an approximation to the
%halos formation process
%We compared
%compared to their results
%our $n=-1$ halos are less centrally concentrated while the halos $n=0$
%have value of $a$ intermediate between that of Navarro et al. (1997)
%and that of Cole and Lacey (1996). For $n=-1$
%the relation between
%scale radius and halo mass is steeper in our model with respect
%to all the others model while for $n=0$ it is comparable with that
%of Navarro et al. (1997). \\

\section{Conclusions}

In this paper we have developed an improved version of the HS model
to study the structure of the dark matter halos. We assumed that
the initial density profile is given by the average profile given by BBKS
and, solving the spherical collapse model, we obtained the virialized density
profile. Our results can be summarized as follows: 
\begin{itemize} 
\item a) Differently from HS's (1985) model the density profiles
are not power-laws but have a logarithmic slope that increase from the
inner halo to its outer parts. In the outer parts of the halo, the
density profiles are steeper than that found by HS
and are consistent with $\rho \propto r^{-3}$, while in the inner part of the
halo we find $\rho \propto \sim r^{-1}$. 
%halo we find $\rho \propto r^{-0.95}$ for $n=-1$ and
%$\rho \propto r^{-1}$ for $n=0$.
The analytic model proposed by
Navarro et al. (1995) is a good fit to the halo profiles. 
\item b) The radius, $a$, at which the slope equals $-2$ is a function
of the mass of the
halo and of the spectral index $n$. 
Lower mass halos are more centrally concentrated than the higher ones.
For a given mass $M$,
halos having larger values of $n$ have denser cores. 
\item c) The good agreement of our (spherical simmetric) model
with several N-body simulations lead us to think, in agreement with
Huss' et al. (1998) paper, that the role
of merging in the formation of halos is not as crucial as generally
believed. 
\end{itemize}

%As shown $a$ increase for increasing $M$. In Fig. 3 $a$ we show the
%variation of $a$ versus $n$. The density profiles are denser
%for larger values of $n$. Finally in Fig. 4, 5 we plot the 
%In Fig. 2 we plot  
%The....line shows the value of the slope in
%the hypothesys that $\delta(r)= \nu \xi(r)/\xi(0)^{1/2}$ and that the
%peak is very high, $\nu>>1$, as done by Hoffman and Shaham (1985). As shown
%we obtain the same profile of the quoted authors. The...line shows the
%logarithmic slope of the density profile at $1 h^{-1} Mpc$
%calculated using BBKS density profile,
%for $\nu=3$. As shown the slope lays between that of
%Hoffman \& Shaham (1985) and that
%of Sheth \& Jain (1996), and is almost coincident
%with that of Shet \& Jain (1985) for $n>-1$.
%%%As expected density profiles
%
%are steeper than Hoffmann \& Shaham's. When we repeat the calculation
%for $nu=2$ we find that an even steeper behaviour of the profile. \\
%In Fig. 2 we plot the variation of the scale parameters $a$ versus mass
%$M$. As shown $a$ increase for increasing $M$. In Fig. 3 $a$ we show the
%variation of $a$ versus $n$. The density profiles are denser
%for larger values of $n$. Finally in Fig. 4, 5 we plot the 

%\begin{flushleft}
%{\it Acknowledgements}
%\end{flushleft}
% We are grateful to .......................for 
%stimulating discussions during the period in which this work was performed.\\
%%%This work was performed in the framework of a MURST ex-$60 \%$ 1998 project.

%

\end{document}